 \documentclass[12pt]{article}
 \usepackage{epsfig}
 \usepackage{amsmath}
 \usepackage{hhline}
 \usepackage{amssymb}
 \usepackage{times}
 \usepackage{cite}
 \usepackage{rotating}
 \usepackage{color}
 \definecolor{darkgreen}{rgb}{0,0.5,0}
 \definecolor{DeepPink3}{rgb}{0.803922,0.062745,0.462745}
 \usepackage[]{lineno}

 \newlength{\dinwidth}
 \newlength{\dinmargin}
\begin{document}  
 \newcommand{\pom}{{I\!\!P}}
 \newcommand{\reg}{{I\!\!R}}
 \newcommand{\slowpi}{\pi_{\mathit{slow}}}
 \newcommand{\fiidiii}{F_2^{D(3)}}
 \newcommand{\fiidiiiarg}{\fiidiii\,(\beta,\,Q^2,\,x)}
 \newcommand{\n}{1.19\pm 0.06 (stat.) \pm0.07 (syst.)}
 \newcommand{\nz}{1.30\pm 0.08 (stat.)^{+0.08}_{-0.14} (syst.)}
 \newcommand{\fiidiiiful}{F_2^{D(4)}\,(\beta,\,Q^2,\,x,\,t)}
 \newcommand{\fiipom}{\tilde F_2^D}
 \newcommand{\ALPHA}{1.10\pm0.03 (stat.) \pm0.04 (syst.)}
 \newcommand{\ALPHAZ}{1.15\pm0.04 (stat.)^{+0.04}_{-0.07} (syst.)}
 \newcommand{\fiipomarg}{\fiipom\,(\beta,\,Q^2)}
 \newcommand{\pomflux}{f_{\pom / p}}
 \newcommand{\nxpom}{1.19\pm 0.06 (stat.) \pm0.07 (syst.)}
 \newcommand {\gapprox}
    {\raisebox{-0.7ex}{$\stackrel {\textstyle>}{\sim}$}}
 \newcommand {\lapprox}
    {\raisebox{-0.7ex}{$\stackrel {\textstyle<}{\sim}$}}
 \def\gsim{\,\lower.25ex\hbox{$\scriptstyle\sim$}\kern-1.30ex%
 \raise 0.55ex\hbox{$\scriptstyle >$}\,}
 \def\lsim{\,\lower.25ex\hbox{$\scriptstyle\sim$}\kern-1.30ex%
 \raise 0.55ex\hbox{$\scriptstyle <$}\,}
 \newcommand{\pomfluxarg}{f_{\pom / p}\,(x_\pom)}
 \newcommand{\dsf}{\mbox{$F_2^{D(3)}$}}
 \newcommand{\dsfva}{\mbox{$F_2^{D(3)}(\beta,Q^2,x_{I\!\!P})$}}
 \newcommand{\dsfvb}{\mbox{$F_2^{D(3)}(\beta,Q^2,x)$}}
 \newcommand{\dsfpom}{$F_2^{I\!\!P}$}
 \newcommand{\gap}{\stackrel{>}{\sim}}
 \newcommand{\lap}{\stackrel{<}{\sim}}
 \newcommand{\fem}{$F_2^{em}$}
 \newcommand{\tsnmp}{$\tilde{\sigma}_{NC}(e^{\mp})$}
 \newcommand{\tsnm}{$\tilde{\sigma}_{NC}(e^-)$}
 \newcommand{\tsnp}{$\tilde{\sigma}_{NC}(e^+)$}
 \newcommand{\st}{$\star$}
 \newcommand{\sst}{$\star \star$}
 \newcommand{\ssst}{$\star \star \star$}
 \newcommand{\sssst}{$\star \star \star \star$}
 \newcommand{\tw}{\theta_W}
 \newcommand{\sw}{\sin{\theta_W}}
 \newcommand{\cw}{\cos{\theta_W}}
 \newcommand{\sww}{\sin^2{\theta_W}}
 \newcommand{\cww}{\cos^2{\theta_W}}
 \newcommand{\trm}{m_{\perp}}
 \newcommand{\trp}{p_{\perp}}
 \newcommand{\trmm}{m_{\perp}^2}
 \newcommand{\trpp}{p_{\perp}^2}
 \newcommand{\alp}{\alpha_s}

 \newcommand{\alps}{\alpha_s}
 \newcommand{\sqrts}{$\sqrt{s}$}
 \newcommand{\LO}{$O(\alpha_s^0)$}
 \newcommand{\Oa}{$O(\alpha_s)$}
 \newcommand{\Oaa}{$O(\alpha_s^2)$}
 \newcommand{\PT}{p_{\perp}}
 \newcommand{\JPSI}{J/\psi}
 \newcommand{\sh}{\hat{s}}
 \newcommand{\uh}{\hat{u}}
 \newcommand{\MP}{m_{J/\psi}}
 \newcommand{\PO}{I\!\!P}
 \newcommand{\xbj}{x}
 \newcommand{\xpom}{x_{\PO}}
 \newcommand{\ttbs}{\char'134}
 \newcommand{\xpomlo}{3\times10^{-4}}  
 \newcommand{\xpomup}{0.05}  
 \newcommand{\dgr}{^\circ}
 \newcommand{\pbarnt}{\,\mbox{{\rm pb$^{-1}$}}}
 \newcommand{\gev}{\,\mbox{GeV}}
 \newcommand{\WBoson}{\mbox{$W$}}
 \newcommand{\fbarn}{\,\mbox{{\rm fb}}}
 \newcommand{\fbarnt}{\,\mbox{{\rm fb$^{-1}$}}}
 \newcommand{\dsdx}[1]{$d\sigma\!/\!d #1\,$}
 \newcommand{\eV}{\mbox{e\hspace{-0.08em}V}}
 %
 %
 \newcommand{\qsq}{\ensuremath{Q^2} }
 \newcommand{\gevsq}{\ensuremath{\mathrm{GeV}^2} }
 \newcommand{\et}{\ensuremath{E_t^*} }
 \newcommand{\rap}{\ensuremath{\eta^*} }
 \newcommand{\gp}{\ensuremath{\gamma^*}p }
 \newcommand{\dsiget}{\ensuremath{{\rm d}\sigma_{ep}/{\rm d}E_t^*} }
 \newcommand{\dsigrap}{\ensuremath{{\rm d}\sigma_{ep}/{\rm d}\eta^*} }

 \newcommand{\dstar}{\ensuremath{D^*}}
 \newcommand{\dstarp}{\ensuremath{D^{*+}}}
 \newcommand{\dstarm}{\ensuremath{D^{*-}}}
 \newcommand{\dstarpm}{\ensuremath{D^{*\pm}}}
 \newcommand{\zDs}{\ensuremath{z(\dstar )}}
 \newcommand{\Wgp}{\ensuremath{W_{\gamma p}}}
 \newcommand{\ptds}{\ensuremath{p_t(\dstar )}}
 \newcommand{\etads}{\ensuremath{\eta(\dstar )}}
 \newcommand{\ptj}{\ensuremath{p_t(\mbox{jet})}}
 \newcommand{\ptjn}[1]{\ensuremath{p_t(\mbox{jet$_{#1}$})}}
 \newcommand{\etaj}{\ensuremath{\eta(\mbox{jet})}}
 \newcommand{\detadsj}{\ensuremath{\eta(\dstar )\, \mbox{-}\, \etaj}}

 \def\Journal#1#2#3#4{{#1} {\bf #2} (#3) #4}
 \def\NCA{\em Nuovo Cimento}
 \def\NIM{\em Nucl. Instrum. Methods}
 \def\NIMA{{\em Nucl. Instrum. Methods} {\bf A}}
 \def\NPB{{\em Nucl. Phys.}   {\bf B}}
 \def\PLB{{\em Phys. Lett.}   {\bf B}}
 \def\PRL{\em Phys. Rev. Lett.}
 \def\PRD{{\em Phys. Rev.}    {\bf D}}
 \def\ZPC{{\em Z. Phys.}      {\bf C}}
 \def\EJC{{\em Eur. Phys. J.} {\bf C}}
 \def\CPC{\em Comp. Phys. Commun.}

 \begin{titlepage}

 \noindent
 DESY $08-172$ \hspace{10 cm}                      ISSN $0418-9833$\\
 February $2009$
 \noindent

 \vspace{2cm}
 \begin{center}
 \begin{Large}

 {\bf Inclusive Photoproduction of \boldmath{$\bf\rho^0$},
  \boldmath{$\bf K^{*0}$} and \boldmath{$\bf \phi$} Mesons 
 at HERA \\}

 \vspace{2cm}

 H1 Collaboration

 \end{Large}
 \end{center}

 \vspace{2cm}

 \begin{abstract}
 \hspace{-0.55 cm}Inclusive non-diffractive photoproduction of $\rho(770)^0$, 
 $K^*(892)^0$ and $\phi(1020)$ mesons is investigated with the H1 detector in $ep$
 collisions at HERA. The corresponding average $\gamma p$ 
 centre-of-mass energy is $210$ GeV. The mesons are measured in
 the transverse momentum range $0.5<p_T<7$ GeV and the rapidity range 
 $|y_{lab}|<1$. Differential cross sections are presented as a function of 
 transverse momentum and rapidity, and are compared to the predictions 
 of hadroproduction models.
 \end{abstract}

 \vspace{1.5cm}

 \begin{center}
Accepted by \PLB \;\; 
 \end{center}

 \end{titlepage}

 %
 %
 %
 \begin{flushleft}

F.D.~Aaron$^{5,49}$,           
C.~Alexa$^{5}$,                
V.~Andreev$^{25}$,             
B.~Antunovic$^{11}$,           
S.~Aplin$^{11}$,               
A.~Asmone$^{33}$,              
A.~Astvatsatourov$^{4}$,       
S.~Backovic$^{30}$,            
A.~Baghdasaryan$^{38}$,        
E.~Barrelet$^{29}$
% ...

\bigskip{\it
% ...
$ ^{4}$ Inter-University Institute for High Energies ULB-VUB, Brussels;
          Universiteit Antwerpen, Antwerpen; Belgium$^{ c}$ \\
 $ ^{5}$ National Institute for Physics and Nuclear Engineering (NIPNE) ,
          Bucharest, Romania \\
% ...
$ ^{11}$ DESY, Hamburg, Germany \\
% ...
$ ^{25}$ Lebedev Physical Institute, Moscow, Russia$^{ e}$ \\
% ...
$ ^{29}$ LPNHE, Universit\'{e}s Paris VI and VII, IN2P3-CNRS,
          Paris, France \\
 $ ^{30}$ Faculty of Science, University of Montenegro,
          Podgorica, Montenegro$^{ e}$ \\
% ...
$ ^{33}$ Dipartimento di Fisica Universit\`a di Roma Tre
          and INFN Roma~3, Roma, Italy \\
% ...
$ ^{38}$ Yerevan Physics Institute, Yerevan, Armenia \\
% ...
$ ^{49}$ Also at Faculty of Physics, University of Bucharest,
          Bucharest, Romania \\
% ...
$ ^c$ Supported by FNRS-FWO-Vlaanderen, IISN-IIKW and IWT
      and  by Interuniversity
Attraction Poles Programme,
      Belgian Science Policy \\
% ...
$ ^e$ Supported by the Deutsche Forschungsgemeinschaft \\
% ...
}
\end{flushleft}
%

\newpage

\section{Introduction}

High energy particle collisions, which give rise to large multiplicities
of produced hadrons, 
provide an opportunity to study the hadronisation process, whereby the quarks 
and gluons produced in the initial interaction become colourless hadrons. 
Since most of these hadrons are produced at low values of transverse momentum, 
perturbative quantum chromodynamics (pQCD) is not applicable
to this process, which is described instead using phenomenological models, 
the most successful of which  
are the string~\cite{string} and the cluster~\cite{cluster} fragmentation models. 
These can provide a reasonable description of the hadronisation process 
provided the many free parameters they contain are tuned to the data.

The production of long-lived hadrons and resonances at high energies has been 
studied in detail in electron-positron ($e^+e^-$) collisions at LEP 
using $Z^0$ decays~\cite{delphireview}.
Measurements in high energy hadronic interactions have so far been restricted
to long-lived hadrons and hadrons containing heavy quarks. 
Recently, the production of the hadronic resonances $\rho(770)^0$, $K^*(892)^0$ 
and $\phi(1020)$
has been measured in heavy-ion and proton-proton ($pp$) collisions at RHIC~\cite{STAR}. 
The electron-proton ($ep$) collider HERA allows the study of particle production
in quasi-real photon-proton ($\gamma p$) collisions.
The comparison of RHIC and HERA results is of particular interest, since
the nuclear density at HERA is much lower than that at RHIC
while the $\gamma p$ and nucleon-nucleon collision energies are similar.

 In this paper, measurements of the inclusive non-diffractive photoproduction of
the resonances $\rho(770)^0$, $K^*(892)^0$ and $\phi(1020)$ at HERA 
are presented for the first time. The measurements are based on
the data recorded with the H1 detector during the year 2000,
when positrons of energy $27.6$ GeV collided with $920$ GeV
protons at an $ep$ centre-of-mass energy of $319$~GeV,
providing on average a $\gamma p$ centre-of-mass
energy of $\langle W\rangle =210$~GeV.
The data correspond to an integrated luminosity of ${\cal L}=36.5$~pb$^{-1}$.

\section{Phenomenology and Monte Carlo Simulation}
The H1 coordinate system has as its origin the position of the nominal 
interaction vertex.
The outgoing proton beam direction defines the positive $z$-axis and is also referred 
to as the ``forward" direction.
The polar angle $\theta$ is defined with respect to this direction. The pseudorapidity
is given by $\eta_{lab}=-\ln(\tan(\theta/2))$. 
The laboratory frame rapidity $y_{lab}$ of a particle 
with energy $E$ and longitudinal momentum $p_z$ is given by 
$y_{lab}=0.5\ln[(E+p_z)/(E-p_z)]$.

The invariant differential cross section for meson production 
can be expressed as a function of
the meson's transverse momentum $p_T$ and its rapidity $y_{lab}$, 
assuming azimuthal symmetry.
Hadrons produced in hadronic collisions are approximately uniformly
distributed in the central rapidity range,
while their transverse momentum spectra fall steeply with increasing $p_T$.
It is convenient to parametrise the invariant differential cross section 
of the produced hadrons with a power law distribution,
\begin{equation}
\frac{1}{\pi}\,\frac{d^2\sigma^{\gamma p}}{dp_T^2\,dy_{lab}} =
\frac{A}{(E_{T_0}+E_{T}^{kin})^n}\; , 
\label{pawer}
\end{equation}
where $E_T^{kin} = \sqrt{m_0^2+p_T^2}-m_0$ is the transverse kinetic energy, 
$m_0$ is the nominal resonance mass, $A$ is a normalisation factor 
independent of $p_T$ and $E_{T_0}$ a free parameter.
When $E_T^{kin}\lesssim E_{T_0}$, the power law 
function~(\ref{pawer}) behaves like a Boltzmann distribution 
$\exp(-E_T^{kin}/T)$, with $T=E_{T_0}/n$.
This exponential behaviour of hadronic spectra follows 
from a thermodynamic model of hadroproduction~\cite{therm}. 
In this framework, the parameter $T$ plays the role of the temperature
at which hadronisation takes place.
At high $E_T^{kin}$, 
the power law originates from a convolution of 
the parton densities of the colliding particles with the 
cross sections of parton-parton interactions. 
The normalisation 
coefficient $A$ is related to the single differential 
cross section $d\sigma/dy_{lab}$ obtained after the integrating 
equation~(\ref{pawer}) over $p_T^2$:
\begin{equation}
A = \frac{d\sigma}{dy_{lab}}\, \frac{(n-1)(n-2)(E_{T_0})^{n-1}}{2\pi (E_{T_0}+(n-2)m_0)}
\; . 
\label{power2}
\end{equation}
Monte Carlo calculations 
are used both to correct the data and in comparisons with the measurements.
Direct and resolved photoproduction
events are simulated using the PYTHIA~\cite{PYTHIA} and 
the PHOJET~\cite{PHOJET} Monte Carlo generators.  
In both cases, the hadronisation is based 
on the string fragmentation model~\cite{JETSET}.
For data corrections, the parameter settings obtained by the ALEPH
collaboration~\cite{BEC} are used for the fragmentation of partons.
The effects of Bose-Einstein correlations (BEC) on the invariant mass spectra 
of like-sign and unlike-sign pion pairs are included using a Gaussian 
parametrisation of the correlation function~\cite{BEC}. 
The photoproduction events generated using PYTHIA and PHOJET 
are passed through the simulation
of the H1 detector based on GEANT~\cite{GEANT} and through the same
reconstruction and analysis chain as used for the data.


\section{Experimental Conditions}
\subsection{H1 Detector}
The H1 detector is described in detail elsewhere \cite{Abt:1997xv}.
A brief account of the components that are most relevant to the
present analysis is given here. 

The $ep$ interaction region is surrounded by two large concentric drift
chambers (CJCs), operated inside a $1.16$~T solenoidal magnetic field. Charged 
particles are measured in the pseudorapidity range $-1.5<\eta_{lab}<1.5$ with a transverse 
momentum resolution of $\sigma_{p_T}/p_T\approx0.005\cdot p_T/$GeV$\; \oplus \; 0.015$~\cite{ptsigma}.
The specific energy loss $dE/dx$ of the charged particles is measured
in this detector with a relative resolution of $7.5\%$ 
for a minimum ionising track~\cite{dedx}. 

A finely segmented electromagnetic and hadronic liquid argon 
calorimeter (LAr) covers the range $-1.5<\eta_{lab}<3.4$. The energy
resolution of this calorimeter is $\sigma(E)/E = 0.11/\sqrt{E/\text{GeV}}$ for
electromagnetic showers and $\sigma(E)/E = 0.50/\sqrt{E/\text{GeV}}$ for
hadrons as measured in test beams~\cite{lar}. 

Photoproduction events are selected with a crystal \u{C}erenkov
calorimeter (positron tagger) located close to the beam pipe 
at $z=-33.4$ m, which measures the energy
deposited by positrons scattered at angles of less than $5$ mrad.
Another \u{C}erenkov calorimeter, located at $z=-103$ m (photon tagger), is used
to determine the luminosity by measuring the rate of photons
emitted in the Bethe-Heitler process $ep\to ep\gamma$. 

\subsection{Event Selection}

Photoproduction events are selected by a trigger which requires
a scattered positron to be measured in the positron tagger, 
an event vertex determined from charged tracks and
three or more charged tracks reconstructed in the CJCs, each with transverse 
momentum $p_T>0.4$~GeV. 
The photon virtuality $Q^2$ is smaller than $0.01$~GeV$^2$,
due to the positron tagger acceptance.
The photon energy is determined from the difference between
the positron beam energy and the energy measured in the 
positron tagger.

In order to reduce the non-$ep$ background 
and to ensure good reconstruction of the event kinematics,
the following criteria are applied:

\begin{itemize}
\item{Events are selected if the reconstructed $\gamma p$ centre-of-mass 
energy lies within the interval $174<W<256$ GeV for which good positron detection 
efficiency is established. This corresponds to an average $\gamma p$ 
centre-of-mass energy of $ \langle W\rangle = 210$ GeV.} 
\item{Events are rejected if a photon with energy $E_\gamma>2$ GeV
is detected in the photon tagger. This suppresses the background arising from 
random coincidences of Bethe-Heitler events in the positron tagger with
beam-gas interactions in the main H1 detector.} 
\item{Events are selected if the $z$ coordinate of the event vertex,
reconstructed using the CJCs, lies within $35$ cm of the mean position
for $ep$ interactions.}
\end{itemize}

Background from elastic and diffractive events is suppressed by the above
trigger requirements. 
To further reduce the contribution of diffractive processes, 
the presence of an energy deposit of at least $500$ MeV 
is required in the forward region of the LAr, defined
by $2.03<\eta_{lab}<3.26$. Monte Carlo studies show that, with 
this requirement, less than $1\%$ of the final event sample consists of 
diffractive events with $X_\pom<0.05$, where $X_\pom=M_X^2/W^2$ and 
$M_X$ is the invariant mass of the diffractive system.

In total, about $1.8\times10^6$ events satisfy the above selection criteria.

\subsection{Selection of \boldmath{$\bf\rho(770)^0$},
 \boldmath{$\bf K^*(892)^0$} and \boldmath{$\bf \phi(1020)$} Mesons}
The mesons are identified using the $\rho(770)^0\to\pi^+\pi^-$,
$K^*(892)^0\to K^+\pi^-$ or $\overline{K}^*(892)^0\to K^-\pi^+$ 
and $\phi(1020)\to K^+K^-$ decays\footnote{
In the following, the notation $K^{*0}$ is used to refer 
to both the $K^{*0}$ and $\overline{K}^{*0}$ mesons unless 
explicitly stated otherwise.}. Charged tracks reconstructed in the CJCs with
$p_T>0.15~$GeV and pseudorapidity $|\eta_{lab}|<1.5$ are considered as
charged pion or kaon candidates. 
Since most of the charged particles in $ep$ collisions are pions, 
no attempt to identify pions is made, while
identification criteria for charged kaons are applied
for the extraction of the $K^{*0}$ and $\phi$ signals.
This is done by measuring the momentum-dependent specific
energy loss $dE/dx$ in the CJCs.
This method gives a significant improvement in the signal-to-background 
ratio for low $p_T$ mesons, $p_T<1.5$~GeV, where the $dE/dx$ 
resolution allows good particle identification. 
For high $p_T$ mesons, $p_T>1.5$~GeV, the $dE/dx$ method is 
inefficient and therefore particle identification 
criteria are not applied. Such tracks are considered as both 
pion and kaon candidates and their four-momenta are determined from
the track measurements using the corresponding mass hypothesis~\cite{pdg}.
Vector meson candidates are reconstructed from these four-momenta.
The kinematic range for the
reconstructed neutral mesons is restricted to $|y_{lab}|<1$ and $p_T>0.5$~GeV.

To extract the $\rho^0$, $K^{*0}$ and $\phi$ signals, 
the distributions of respective invariant masses of their decay products, 
$m_{\pi^+\pi^-}$, $m_{K^\pm\pi^\mp}$ and $m_{K^+ K^-}$, are fitted using 
a function composed of three parts:
\begin{equation}
F(m)=B(m)+\sum R(m)+\sum S(m).
 \label{invfit}
\end{equation}
The terms correspond to contributions from the combinatorial background, $B(m)$, 
from reflections which result from decays other than the signal
  under consideration, $R(m)$, and from the relevant signal, $S(m)$, respectively.

The combinatorial background function is taken to be:
$$
B(m) = (a_0+a_1m+a_2m^2+a_3m^3)\cdot B^0(m)\, ,
$$
where $a_0$, $a_1$, $a_2$ and $a_3$ are free parameters,
and $B^0(m)$ is the invariant mass distribution of the like-sign charged particle combinations:
$\pi^\pm\pi^\pm$ for the $\rho^0$ and $K^\pm\pi^\pm$ for the $K^{*0}$.
The shape of the combinatorial background for $\phi$ is 
described by the following function:
$$
B(m) = b_1(m^2-4m_K^2)^{b_2} e^{-b_3 m}\, ,
$$
where $b_1$, $b_2$ and $b_3$ are free parameters and $m_K$ is the kaon mass.

The second term, $\sum R(m)$, in~(\ref{invfit}) represents the sum of the reflections;
for example, charged particles from the decay $K^{*0}\to K^\pm \pi^\mp$ with the kaon 
misidentified as a charged pion will give  rise to structure in the $m_{\pi^+\pi^-}$ 
spectrum and must be taken into account as a separate contribution. 
In addition, there are two other contributions to the $m_{\pi^+\pi^-}$ spectrum 
in the mass region of interest. These arise
from the decays $\omega(782)\rightarrow \pi^+ \pi^-$ and 
$\omega(782)\rightarrow \pi^+ \pi^-\pi^0$
in which the $\pi^0$ is not observed.
%
For the $\omega(782)$ meson, the production rate relative to that
of the $\rho^0$ is varied within the range $1.0\pm0.2$,
which is consistent with measurements of the $\omega(782)/\rho^0$ ratio in
hadronic collisions~\cite{reflection_ISR} and 
in $Z^0$ boson decays~\cite{reflection_LEP}.
The $\omega(782)$ branching ratios are taken from ~\cite{pdg}.
The five major reflections in the $m_{K^\pm\pi^\mp}$ spectrum 
are due to: the decay 
$\rho^0\rightarrow \pi^+ \pi^-$ with the $\pi^+$ or $\pi^-$ misidentified as 
a charged kaon; the decays 
$\omega(782)\rightarrow \pi^+ \pi^-$ and 
$\omega(782)\rightarrow \pi^+ \pi^-\pi^0$
with the $\pi^0$ not observed and with one of the $\pi^+$ or $\pi^-$ 
mesons misidentified as a charged kaon; the decay 
$\phi\rightarrow K^+K^-$ with one of the
kaons misidentified as a charged pion and 
a self-reflection from the $K^{*0}$, where the pion and kaon
are interchanged.   
For the $m_{K^+K^-}$ spectrum, there are no reflections from known resonances
in the invariant mass region of interest.
Therefore, the shapes of the reflections are taken from Monte Carlo
calculations. 
The contribution of the reflections 
from the $\rho^0$, $K^{*0}$ and $\phi$ mesons
is tied to the production rates determined in this analysis 
and is therefore iteratively calculated.

The function $S(m)$ used to describe the signal in~(\ref{invfit}) 
is a convolution of a relativistic Breit-Wigner function $BW(m)$
and a detector resolution function $r(m,m')$.
The relativistic Breit-Wigner function 
\begin{equation}
 BW(m)=A_0\,\frac{m_0\,m\,\Gamma(m)}{(m^2-m_{0}^{2})^2+m_{0}^{2}\,\Gamma^{2}(m)}\;,
 \label{relbw}
\end{equation}
is used with
$$\Gamma(m)=\Gamma_{0}\left(\frac{q}{q_0}\right)^{2l+1}\frac{m_0}{m}\;,$$
where $A_0$ is a normalisation factor, $\Gamma_{0}$ is the resonance width,
$l=1$ for vector mesons, $m_0$ is the resonance mass,
$q$ is the momentum of the decay products in the rest frame
of the parent meson, and $q_0$ is their momentum at $m=m_0$.
The cross sections cited in this paper assume that the meson signal is
defined as the integral of the relativistic Breit-Wigner function~(\ref{relbw})
in the region $\pm2.5\Gamma_0$ around the mass $m_0$.
Monte Carlo studies show that a non-relativistic Breit-Wigner function with width $\Gamma_{res}$
provides a good description of the detector resolution function:
\begin{equation}
 r(m,m') = \frac{1}{2\pi}\,\frac{\Gamma_{res}}{(m-m')^2+(\Gamma_{res}/2)^2}\;.
 \label{resol}
\end{equation}

For the $K^{*0}$ analysis, the resolution parameter is determined 
from Monte Carlo with $\Gamma_{res}=12$~MeV. 
It is small compared to the width of the $K^{*0}$ 
meson ($50.3\pm0.6$~MeV)~\cite{pdg},
leading only to a small change in the shape of the resonance. 
For the $\phi$, $\Gamma_{res}$
is comparable to the 
width of the $\phi$ meson ($\Gamma_0=4.26\pm0.05$~MeV)~\cite{pdg}. 
As a result, the shape of the $\phi$ signal is 
significantly changed, and hence the detector resolution $\Gamma_{res}$ 
is taken as a free parameter in the fit. It is found to vary from $3.4$~MeV  
to $6.0$~MeV, increasing with the $p_T$ of the $\phi$ meson.

\begin{figure}[ht]
\center
\setlength{\unitlength}{1cm}
\hspace*{1.5cm}
\begin{picture}(15.0,10.0)
\epsfig{file=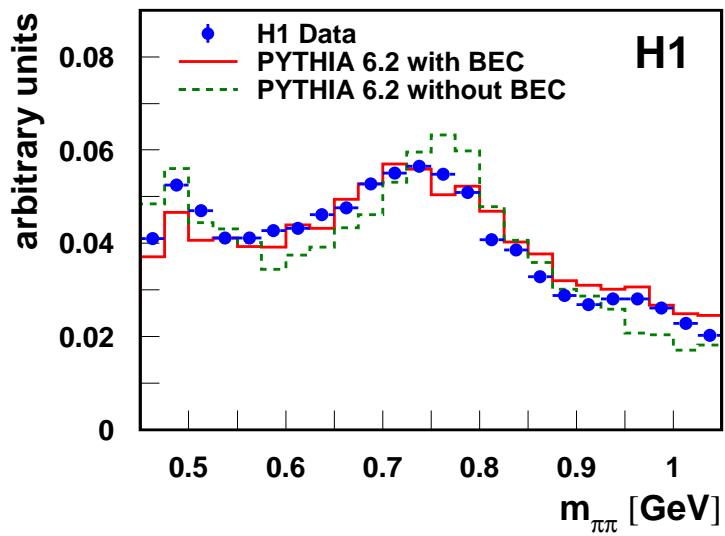,
        bburx=350,bbury=570,bbllx=20,bblly=300,
        width=10.0cm}
\end{picture}
\caption{
The unlike-sign di-pion mass spectrum after subtracting the like-sign contribution, 
normalised to the total number of entries.
The solid and dashed curves correspond  to the PYTHIA simulation
with and without Bose-Einstein correlations (BEC), respectively.
}
\label{bec}
\end{figure}

For the $\rho^0$ meson, the detector resolution is significantly 
smaller than its width.
However, BEC between the $\rho^0$ decay pions and other pions in the event
strongly distort the $\rho^0$ line shape. 
The BEC plays an important role in broadening the $\rho^0$ 
mass peak and in shifting it towards lower masses.
Similar effects are observed in $pp$ and heavy-ion collisions 
at RHIC~\cite{STAR} and in $e^+e^-$ collisions
at LEP using $Z^0$ decays~\cite{reflection_LEP}. 
It is therefore important to check that the Monte Carlo
model used for the extraction of the cross sections 
describes the di-pion spectra in the data.
The data spectra and
the Monte Carlo simulations with and without BEC
are shown in figure~\ref{bec}.
The Monte Carlo model with BEC is 
in a good agreement with the data in the region of the $\rho^0$ resonance, 
whereas the model without BEC fails to describe the di-pion mass spectrum.

\begin{figure}[ht]
\center
\setlength{\unitlength}{1cm}
\hspace*{1.5cm}
\begin{picture}(15.0,10.0)
\epsfig{file=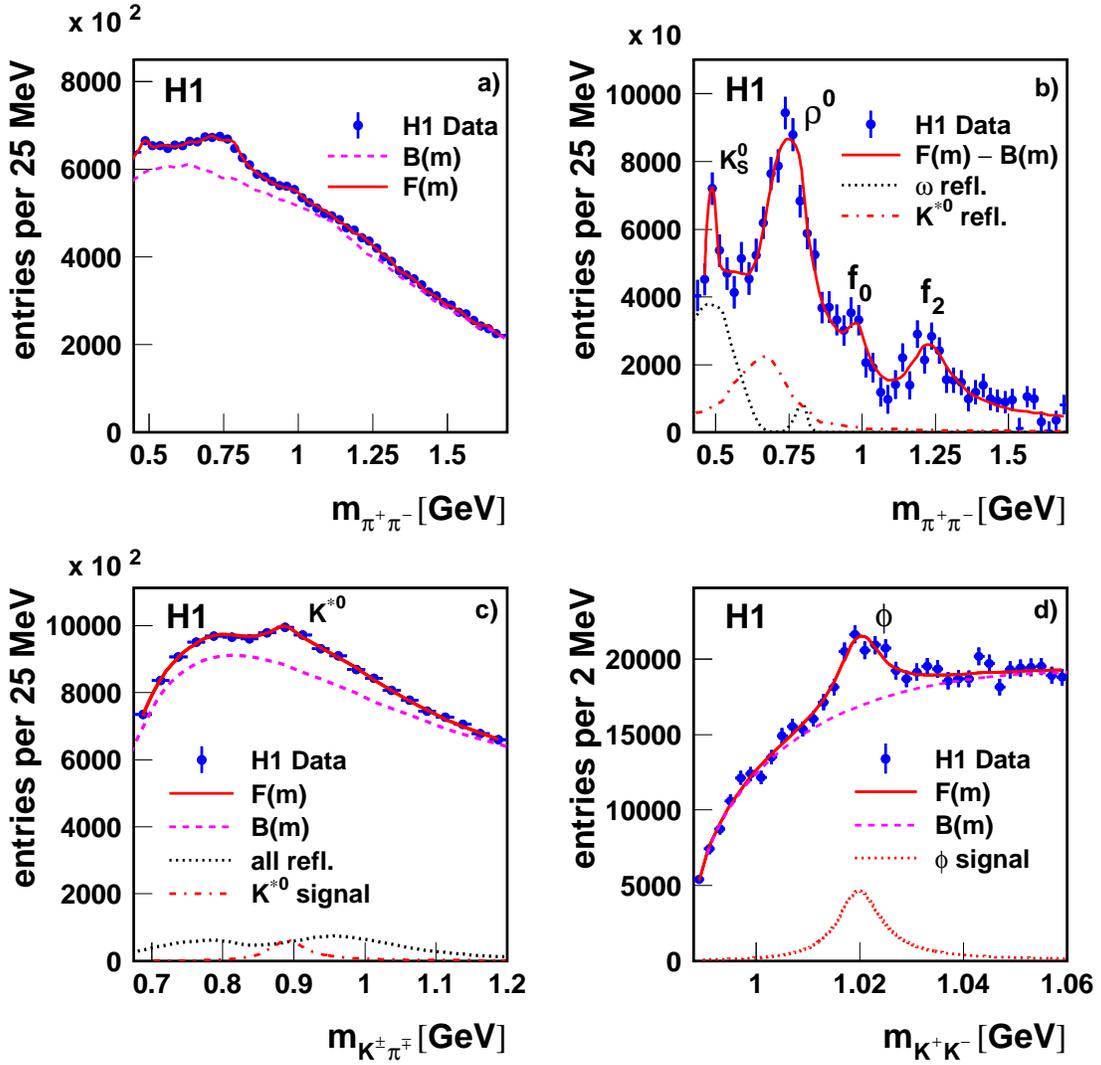,        
        bburx=530,bbury=600,bbllx=50,bblly=80,
        clip=,width=14.0cm,angle=0}
\end{picture}
\caption{The invariant mass spectra for $\pi^+\pi^-$ in $a)$ and $b$), for
  $K^\pm\pi^\mp$ in $c$) and for $K^+K^-$ in~$d$). 
The full curves show the result of the fit; 
the dashed curves correspond to the contribution of the combinatorial background $B(m)$.
In~$b)$, the data and the fit $F(m)$ are shown after subtraction 
of the combinatorial background $B(m)$; the dotted and dash-dotted curves show the
contributions from $\omega$ and $K^*$ reflections, respectively.
In~$c)$, the dotted curve corresponds to the contribution of the reflections and
the dash-dotted curve corresponds to the contribution of the $K^*(892)$ signal.
In~$d)$, the dotted curve corresponds to the contribution of the $\phi(1020)$ signal.
}
\label{fig:signal}
\end{figure}

The results of fitting the function~(\ref{invfit})
to the $m_{\pi^+\pi^-}$ data in the mass range from $0.45$ to 
$1.7~$GeV with the contributions due to the combinatorial background and  
the reflections are shown in figure~\ref{fig:signal}a$)$, and 
after combinatorial background subtraction in figure~\ref{fig:signal}b$)$. 
In this mass range, the signal from the $K^0_S$, $f_0(980)$ and $f_2(1270)$ mesons
is taken into account.
The $K^0_S$ signal is fitted using a Gaussian centred on the
nominal mass and with fixed width.
The relativistic Breit-Wigner function given in equation~(\ref{relbw})
is used for the $f_0(980)$ and $f_2(1270)$ mesons.
In the fit, the resonance masses $m_0$ 
and the yields
are free parameters.
The $\rho^0$ and $f_2(1270)$ widths are fixed to 
the Particle Data Group~\cite{pdg} values 
and the $f_0(980)$ width is fixed to $70$~MeV. 
Due to the small signal and the non-trivial background behaviour, which lead  
to large uncertainties,
cross sections for $f_0(980)$ and $f_2(1270)$ meson production are
not measured here.

The $K^{*0}$ signal is measured 
under the assumption that there is no difference between 
the particle and antiparticle production rates, and
the signal obtained from the $m_{K^\pm \pi^\mp}$ spectrum is divided 
by $2$ to determine the $K^{*0}$ rate in the following. 
The result of fitting the function~(\ref{invfit})
to the $m_{K^\pm\pi^\mp}$ data in the mass range from $0.7$ to
$1.2~$GeV with the contributions due to the combinatorial background and the 
reflections is shown in figure~\ref{fig:signal}c$)$.
In the fit, the $K^{*0}$ width is fixed to the nominal value
while the mass parameter is left free. The result for the $K^{*0}$ mass 
is compatible with the world average~\cite{pdg}.  
 

The result of fitting function~(\ref{invfit})
to the $m_{K^+K^-}$ data in the mass range from $0.99$ to
$1.06~$GeV, together with the background contribution, is shown in 
figure~\ref{fig:signal}d$)$. In the fit, the $\phi$ width, $\Gamma_0$, is fixed 
to the nominal value while the mass is left a 
free parameter and is found to be compatible
with the world average~\cite{pdg}. 

%

\subsection{Cross Section Determination and Systematic Errors}

The invariant differential cross section for $\rho^0$, $K^{*0}$ and 
$\phi$ meson production is measured in the $y_{lab}$ region from
$-1$ to $1$ in seven bins in transverse momentum from $0.5$ to $7$~GeV.
It is calculated according to: 
$$
\frac{1}{\pi}\,\frac{d^2\sigma^{\gamma p}}{dp_T^2\,dy_{lab}} = \frac{N}
{\pi \cdot {\cal L}\cdot BR \cdot \Phi_\gamma\cdot \epsilon 
\cdot \Delta p_T^2\cdot \Delta y_{lab}}
\, , 
$$
where $N$ is the number of mesons from the fit in each bin.
The corresponding bin widths are $\Delta y_{lab}=2$ and 
$\Delta p_T^2 = 2p_T^{bin}\Delta p_T$.
Bin centre corrections based on equation~(\ref{pawer}) 
are applied to define the value of $p_T^{bin}$
at which the differential cross section is measured. 
${\cal L}$ denotes the integrated luminosity and $\epsilon$ the efficiency. 
The branching fractions $BR$ are taken from~\cite{pdg}
and are equal to $1$, $0.67$ and $0.49$ for $\rho^0\to \pi^+\pi^-$, 
$K^{*0}\to K^\pm\pi^\mp$ and $\phi\to K^+K^-$, respectively.
The photon flux $\Phi_\gamma = 0.0127$ is calculated
using the Weizs\"acker-Williams approximation~\cite{Weizs}.

The single differential cross section for $\rho^0$, $K^{*0}$ and
$\phi$ meson production for $p_T>0.5$~GeV is measured in four
bins in rapidity from $-1$ to $1$ according to:
$$
\frac{d\sigma^{\gamma p}}{dy_{lab}} = \frac{N}
{{\cal L}\cdot BR \cdot \Phi_\gamma\cdot \epsilon
\cdot \Delta y_{lab}}
\, .
$$
Here, the bin width is $\Delta y_{lab}=0.5$. 

The fit procedure described in the previous section is repeated to
determine the number of mesons, $N$, in each measurement bin,
calculated as an integral over the signal function~(\ref{relbw})
within $\pm2.5\Gamma_0$ around the meson mass.
Similarly, the total visible cross section for $\rho^0$, $K^{*0}$ and
$\phi$ meson production is measured from the number of
mesons fitted in the range $|y_{lab}|<1$ and $p_T>0.5$~GeV.

The efficiency is given by 
$\epsilon=\epsilon_{rec}\cdot {\cal A}_{etag}\cdot {\cal A}_3\cdot\epsilon_{trig}$.
The reconstruction efficiency for the mesons, $\epsilon_{rec}$, 
includes the geometric acceptance and the efficiency for track 
reconstruction. It is calculated using Monte Carlo data and  
is at least $45\%$ at low $p_T$ 
and rises to about $90\%$ with increasing $p_T$.
For the acceptance determination, the Monte Carlo generators are reweighted to
model the observed $p_T$-dependences.
The average acceptance of the positron tagger, ${\cal A}_{etag}$, is about $50\%$,
as determined in~\cite{H1total}.
The trigger acceptance, ${\cal A}_3$, arises from the requirement that at least 
three tracks are reconstructed
in the CJCs with $p_T>0.4$ GeV. It is determined from Monte Carlo 
simulations with PYTHIA and PHOJET and varies from $50\%$ to $95\%$. 
The trigger efficiency, $\epsilon_{trig}$, is calculated from the data 
using monitor triggers. It is about $90\%$.
The efficiencies and acceptances as
calculated from the PYTHIA and PHOJET simulation are found to be consistent. 
Small residual differences, attributed to different track multiplicity
predictions, 
are taken into account in the systematic uncertainties of the measurement.

The statistical error varies from $7$ to $15\%$ for 
the $\rho^0$, $10$ to $18\%$ for the $K^{*0}$ and 
$13$ to $24\%$ for the $\phi$ meson cross sections.
The systematic errors arise from the uncertainties in the
track reconstruction efficiency ($4\%$) and the trigger efficiency (up to $6\%$),
the variation of the $f_0(980)$ width by $\pm30$~MeV in the $\rho^0$  
fit (up to $7\%$), 
the uncertainties in the $dE/dx$ kaon identification procedure 
($6\%$ for the $K^{*0}$ and 
$12\%$ for the $\phi$) and the luminosity calculation ($2\%$),
the variation of the background shape ($5\%$) and 
the variation of the assumptions about the normalisation
of the contributions from the reflections 
($4\%$ for the $\rho^0$ and up to $15\%$ for the $K^{*0}$).   
The total systematic error varies from $10$ to $12\%$ for the $\rho^0$,
$11$ to $21\%$ for the $K^{*0}$ and $10$ to $17\%$ 
for the $\phi$ meson cross sections.
%
\section{Results and Discussion}


The inclusive non-diffractive photoproduction cross sections for
$\rho^0(770)$, $K^{*0}(892)$ and $\phi(1020)$ mesons in the
kinematic region $Q^2<0.01$ GeV$^2$, $174<W<256$ GeV, and 
for $p_T>0.5$ GeV and $|y_{lab}|<1$ are found to be:
$$
\phantom{} \sigma_{vis}^{\gamma p}(\gamma p \rightarrow \rho^0 X)
\phantom{0}\, = 25600 \pm 1800 \pm 2700 \text{ nb;}
$$
$$
 \sigma_{vis}^{\gamma p}(\gamma p \rightarrow K^{*0} X) 
= \phantom{0}6260 \pm \phantom{0}350 \pm \phantom{0}860 \text{ nb;}
$$
$$
\, \sigma_{vis}^{\gamma p}(\gamma p \rightarrow \phi X) 
\phantom{00}= \phantom{0}2400 \pm \phantom{0}180 \pm \phantom{0}340 \text{ nb.}
$$
The first error is statistical and the second systematic.
Note that the $K^{*0}$ cross section is the sum of 
the particle and antiparticle contributions divided by $2$.

\begin{figure}[ht]
\center
\setlength{\unitlength}{1cm}
\begin{picture}(17.0,16.0)
\epsfig{file=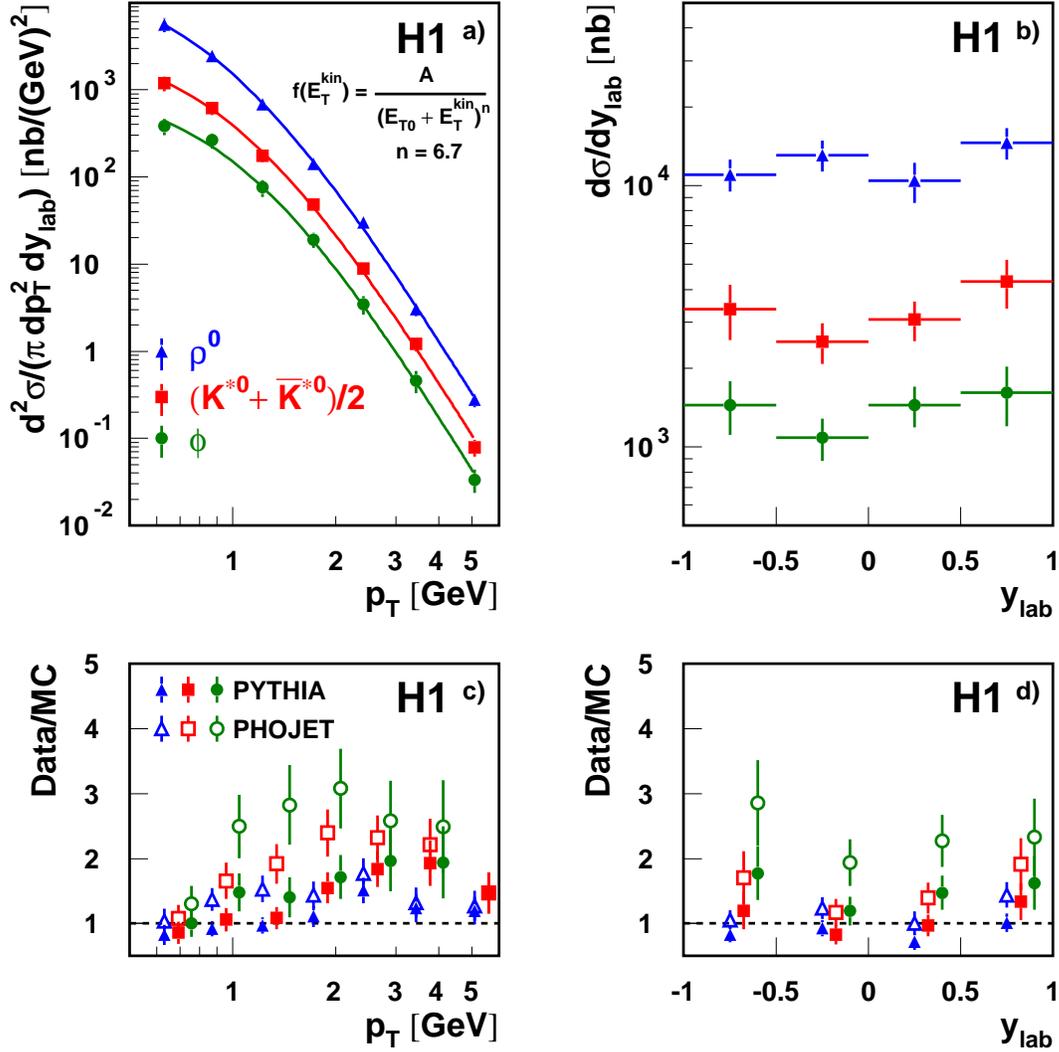,
        bburx=520,bbury=580,bbllx=0,bblly=40,
        clip=,width=15.0cm}
\end{picture}
\caption{The inclusive differential non-diffractive cross sections for
$\rho^0(770)$, $K^{*0}(892)$ and $\phi(1020)$ mesons measured in~$a)$ as a function of
transverse momentum for $|y_{lab}|<1$ and in~$b)$ as a function of 
rapidity for $p_T>0.5$ GeV.
The curves on the figure~$a)$ correspond to the power law,  
equation~(\ref{pawer}), with $n=6.7$.
The ratios of data to Monte Carlo predictions ``Data/MC'' are shown for 
the PYTHIA (full points) and PHOJET (empty
points) simulations as a function of transverse momentum 
for $|y_{lab}|<1$ in~$c)$ and as a function of rapidity for $p_T>0.5$ GeV in~$d)$. 
Statistical and systematic errors are added in quadrature.}
\label{fig:sigma}
\end{figure}

\begin{figure}[ht]
\center
\setlength{\unitlength}{1cm}
\vspace*{4.0cm}
\begin{picture}(15.0,10.0)
\epsfig{file=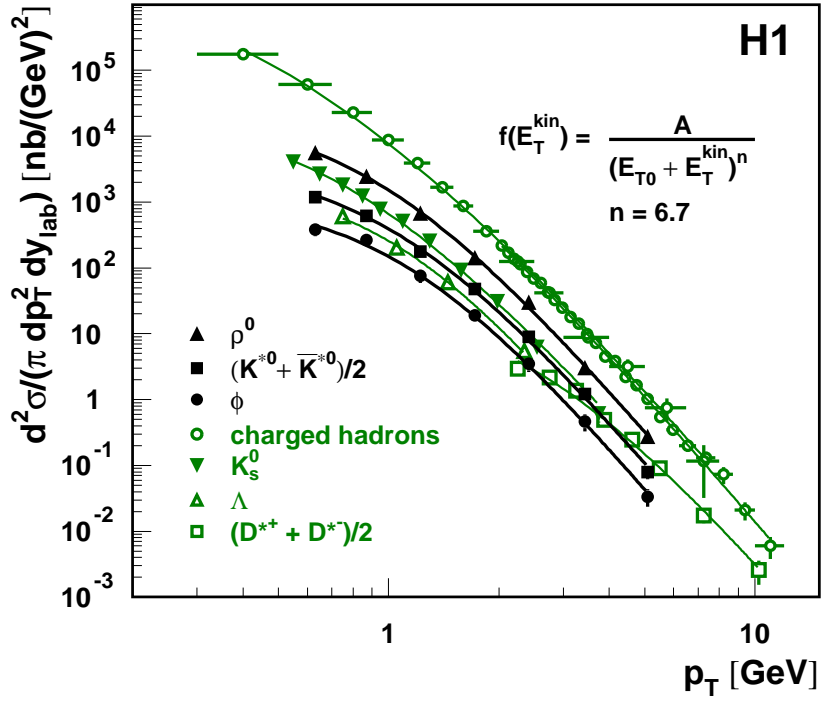,
        bburx=530,bbury=580,bbllx=120,bblly=220,
        clip=,width=12.0cm}
\end{picture}
\caption{
The inclusive invariant differential cross sections as a function of transverse
momentum. The curves show the results of fits to the power law, equation~(\ref{pawer}).
Statistical and systematic errors are added in quadrature.
}
\label{fig:allxsec}
\end{figure}

The differential cross sections for the photoproduction of $\rho^0$, 
$K^{*0}$, and $\phi$ mesons are presented in tables~$1$~and~$2$ and in 
figure~\ref{fig:sigma}. Within the rapidity range of this measurement, 
the resonance production rates are constant as a function of rapidity,
within errors.
The transverse momentum spectra of the $\rho^0$, $K^{*0}$ and $\phi$ mesons 
can be parametrised by function~(\ref{pawer}), where
$d\sigma/dy_{lab}$ in equation~(\ref{power2}) corresponds to the average value of 
the cross section over central rapidities, 
$\langle d\sigma/dy_{lab}\rangle_{|y_{lab}|<1}$.
In the fit, the value of the power $n$ is fixed to be $6.7$, as derived previously  
from measurements of charged particle spectra by the 
H1 collaboration~\cite{H1pions} which gave $n=6.7\pm0.3$. 
The power law distribution, with this value of
$n$, describes 
$K^0_S$ meson, $\Lambda^0$ baryon~\cite{HERA_L0} and $D^{*\pm}$ 
meson production~\cite{HERA_charmed} at HERA,
as is shown in figure~\ref{fig:allxsec}.  
A similar shape of the transverse momentum distribution, 
but with different values of the parameters $n$ and $E_{T_0}$,
was reported for charged particles produced in hadronic 
collisions~\cite{UA1charged}.
The results of the fits of the data to function~(\ref{pawer}) 
are shown in figure~\ref{fig:sigma}a$)$. 
In table~$3$, the parameters of the fit and the average transverse 
kinetic energy $\langle E_T^{kin}\rangle$, the average transverse energy 
$\langle E_T\rangle=\langle E_T^{kin}\rangle +m_0$ and
the average transverse momentum 
$\langle p_T\rangle = \sqrt{\langle E_T\rangle^2-m_0^2}$
derived from~(\ref{pawer}) are presented. 
The errors include the experimental uncertainty on the value of $n$.
Also given are the $\langle p_T\rangle$ values measured at RHIC in $pp$ and
Au-Au collisions~\cite{STAR}
. 

It is interesting to observe that the resonances with different masses,
lifetimes and strangeness content are produced with about the same value
of the average transverse kinetic energy $\langle E_T^{kin}\rangle$. 
This observation supports the
thermodynamic picture of hadronic interactions~\cite{therm}, in which the
primary hadrons are thermalised during the interaction.
The values of $\langle p_T\rangle$ for $\rho^0$, $K^{*0}$ 
and $\phi$ mesons are similar in $\gamma p$ and $pp$
collisions with about the same centre-of-mass energy $\sqrt{s} \approx 200$~GeV, 
while these values are all higher in Au-Au collisions.

The PYTHIA and PHOJET models do not describe the shape of the measured
$p_T$ spectra. Moreover, contrary to the data, the Monte Carlo $p_T$ 
spectra are not described by the power law function~(\ref{pawer}). 
These observations are illustrated in figures~\ref{fig:sigma}c$)$ 
and~\ref{fig:sigma}d$)$.

The measurements in the visible kinematic range of the $\rho^0$, $K^{*0}$ 
and $\phi$ mesons, $p_T>0.5$~GeV and $|y_{lab}|<1$, are extrapolated 
to the full $p_T$ range using the parametrisation~(\ref{pawer}) 
to determine the total inclusive non-diffractive photoproduction cross sections.
The extrapolation factors are of order two.
In the rapidity interval $|y_{lab}|<1$ and integrated over the full $p_T$ range
the following cross section ratios are obtained:
$$ \,R(K^{*0}/\rho^0)
=0.221\pm0.036\, ;$$
$$\quad\, R(\phi/\rho^0)
=0.078\pm0.013\, ; $$
$$\;\, R(\phi/K^{*0})
=0.354\pm0.060\, .$$
The errors are given by the statistical and systematic errors 
added in quadrature. PYTHIA and PHOJET, with the strangeness 
suppression factor $\lambda_s = 0.286$~\cite{BEC},
predict the ratios $0.200$, $0.055$ and $0.277$, respectively,   
which are similar to the measured values, but are all somewhat  lower than these.
 
In table~$4$, 
$R(\phi/K^{*0})$ is compared 
to the corresponding ratios measured by STAR in $pp$
and Au-Au collisions~\cite{STAR} at $\sqrt{s_{NN}}=200$ GeV.
Although the rapidity ranges at the H1 and RHIC experiments 
differ\footnote{The difference in rapidity between the laboratory
frame and the $\gamma p$ frame is about two units at H1.}, 
the resulting ratios for $pp$ and $\gamma p$ interactions are
very close.
However, the corresponding result in Au-Au collisions is observed to be higher.

\section{Conclusions}

First measurements of the inclusive non-diffractive photoproduction of $\rho(770)^0$, 
$K^*(892)^0$ and $\phi(1020)$ mesons at HERA are presented. The differential 
cross sections for the production of these resonances as a function of 
transverse momentum 
are described by a power law distribution while the single differential 
cross sections as a function of rapidity are observed to be flat 
in the visible range.
Despite their different masses,
lifetimes and strangeness content, these resonances  are produced with about the same value
of the average transverse kinetic energy. This observation supports a 
thermodynamic picture of hadronic interactions.

The description of the shape of the $\rho^0$ resonance produced in $\gamma p$
collisions at HERA is improved by taking Bose-Einstein correlations 
into account. A similar effect is observed in
$pp$ and heavy-ion collisions at RHIC and in $e^+e^-$ annihilation at LEP, using $Z^0$
decays. 

The cross section ratios $R(K^{*0}/\rho^0)$, $R(\phi/\rho^0)$ and 
$R(\phi/K^{*0})$ are determined, and $R(\phi/K^{*0})$ is compared to 
results obtained in $pp$ and heavy-ion collisions by the STAR experiment at RHIC.
The ratio $R(\phi/K^{*0})$ measured in $\gamma p$ interactions 
is in agreement with the $pp$ results, while this ratio is observed to be
smaller than the result obtained in Au-Au collisions.

\section*{Acknowledgements}

We are grateful to the HERA machine group whose outstanding
efforts have made this experiment possible. 
We thank the engineers and technicians for their work in constructing and
maintaining the H1 detector, our funding agencies for 
financial support, the
DESY technical staff for continual assistance
and the DESY directorate for support and for the
hospitality which they extend to the non-DESY 
members of the collaboration.

\clearpage
\renewcommand{\arraystretch}{1.4}
\begin{table}
    \begin{center}
     \begin{tabular}{|c|c|c|c|c|}
      \hline
      \multicolumn{2}{|c|}{~} & \multicolumn{3}{|c|}
      {\phantom{$\frac{\frac{\text{\large 0}}{0}}{\frac{\text{\large 0}}{0}}$}
        {\Large $\frac{1}{\pi}\, \frac{d^2\sigma}{dp_T^2\,dy_{lab}}\,$}$[$nb/(GeV)$^2]$}\\
      \hline
      $p_T$ [GeV] & $p_T^{bin}$ &$\rho^0$& $(K^{*0}+\overline{K}^{*0})/2$& $\phi$\\
      \hline
            $[0.5,0.75]$&$0.63$ & $\phantom{.}5610\pm\phantom{.0}870\pm\phantom{.0}590$ & 
            $\phantom{.}1190\pm\phantom{.0}130\pm\phantom{.0}200$ & 
            $\phantom{.00}383\pm\phantom{.000}54\pm\phantom{.000}60$ \\
      \hline
            $[0.75,1.0]$&$0.87$ & $\phantom{.}2440\pm\phantom{.0}180\pm\phantom{.0}260$ & 
            $\phantom{.0}621\pm\phantom{.00}68\pm\phantom{.00}80$ & 
            $\phantom{.00}264\pm\phantom{.000}34\pm\phantom{.000}37$ \\
      \hline
            $[1.0,1.5]$&$1.22$ & \phantom{.0}$680\pm\phantom{.00}55\pm\phantom{.00}70$ & 
            $\phantom{.0}176\pm\phantom{.00}18\pm\phantom{.00}21$ & 
            $\phantom{.000}76\pm\phantom{.000}12\pm\phantom{.000}11$ \\
      \hline
            $[1.5,2.0]$&$1.72$ & $\phantom{.0}142\pm\phantom{.00}15\pm\phantom{.00}15$ & 
            $\phantom{0}48.0\pm\phantom{00}5.2\pm\phantom{00}5.1$ & 
            $\phantom{00}19.1\pm\phantom{000}3.3\pm\phantom{000}1.9$ \\
      \hline
            $[2.0,3.0]$&$2.41$ & $\phantom{0}29.9\pm\phantom{00}2.3\pm\phantom{00}3.1$ & 
            $\phantom{0}8.96\pm\phantom{0}0.90\pm\phantom{0}0.98$ & 
            $\phantom{00}3.48\pm\phantom{00}0.76\pm\phantom{00}0.34$ \\
      \hline
            $[3.0,4.0]$&$3.43$ & $\phantom{0}3.06\pm\phantom{0}0.42\pm\phantom{0}0.33$ & 
            $\phantom{0}1.21\pm\phantom{0}0.17\pm\phantom{0}0.14$ & 
            $\phantom{00}0.46\pm\phantom{00}0.11\pm\phantom{00}0.08$ \\
      \hline
            $[4.0,7.0]$&$5.09$ & $0.276\pm0.037\pm0.033$ & 
            $0.079\pm0.014\pm0.009$ & $0.0335\pm0.0081\pm0.0057$ \\
      \hline
    \end{tabular}
    \end{center}
\caption{
Inclusive non-diffractive photoproduction invariant differential cross sections
$d^2\sigma/\pi\, dp_T^2\,dy_{lab}$ for $\rho(770)^0$, $K^*(892)^0$ and $\phi(1020)$ mesons 
in the rapidity range $|y_{lab}|<1.0$ in bins of $p_T$.
The first error is statistical and the second systematic.
For each bin in $p_T$ the range as well as the bin-centred value $p_T^{bin}$ are given.
}
\end{table}
\begin{table}
    \begin{center}
     \begin{tabular}{|c|c|c|c|}
      \hline &
      \multicolumn{3}{|c|}{$ d\sigma/dy_{lab}$~[$\mu$b]} \\
      \hline
       $y_{lab}$&$\rho^0$& $(K^{*0}+\overline{K}^{*0})/2$& $\phi$\\
      \hline
      $[-1.0,-0.5]$ & $11.0\pm1.0\pm1.2$ & $3.36\pm0.35\pm0.72$ & $1.44\pm0.25\pm0.22$ \\
      \hline
      $[-0.5,0.0]$ & $13.1\pm1.1\pm1.4$ & $2.52\pm0.27\pm0.36$ & $1.08\pm0.12\pm0.16$ \\
      \hline
      $[0.0,0.5]$ & $10.4\pm1.5\pm1.1$ & $3.07\pm0.30\pm0.44$ & $1.44\pm0.13\pm0.22$ \\
      \hline
      $[0.5,1.0]$ & $14.6\pm1.3\pm1.5$ & $4.28\pm0.44\pm0.79$ & $1.61\pm0.33\pm0.25$ \\
      \hline
     \end{tabular}
    \end{center}
\caption{
Inclusive non-diffractive photoproduction single differential 
cross sections $d\sigma/dy_{lab}$
for $\rho(770)^0$, $K^*(892)^0$ and $\phi(1020)$ mesons  
in the transverse momentum range $p_T>0.5$ GeV in bins of $y_{lab}$. 
The first error is statistical and the second systematic.}
\end{table}

\clearpage
\begin{table}
    \begin{center}
     \begin{tabular}{|l|r|c|c|c|}
      \hline
       \multicolumn{2}{|c|}{} &$\rho^0$& $(K^{*0}+\overline{K}^{*0})/2$& $\phi$\\
      \hline
      $\gamma p$ (H1) &$\langle d\sigma/dy_{lab}\rangle_{|y_{lab}|<1}$ [$\mu$b] &$23.6\pm2.7$  &
       $\phantom{.}5.22\pm0.60\phantom{.0}$&$\phantom{.}1.85\pm0.23\phantom{.0}$ \\
      \cline{2-5}
      &$T$ [GeV] &$\phantom{0}0.151\pm0.011$  &$0.166\pm0.012$&$0.170\pm0.012$ \\
      \cline{2-5}
      &$\langle E_T\rangle$ [GeV]&$\phantom{0}1.062\pm0.018 $  &$1.205\pm0.020$&$1.333\pm0.022$ \\
      \cline{2-5}
      &$\langle E_T^{kin}\rangle$ [GeV]&$\phantom{0}0.287\pm0.018$  &$0.313\pm0.020$&$0.314\pm0.022$ \\
      \cline{2-5}
      &$\langle p_T\rangle$ [GeV]&$\phantom{0}0.726\pm0.027$  &$0.810\pm0.030$&$0.860\pm0.035$ \\
      \hline
      \hline
      $pp$ (STAR) &$\langle p_T\rangle_{pp}$ [GeV]&$\phantom{0}0.616\pm0.062$  &$0.81\pm0.14$& $0.82\pm0.03$\\
      \hline
      Au-Au (STAR) &$\langle p_T\rangle_{AuAu}$ [GeV]&$\phantom{00}0.83\pm0.10\phantom{0}$  &$1.08\pm0.14$&$0.97\pm0.02$ \\
      \hline
     \end{tabular}
    \end{center}
\caption{The parameters $\langle d\sigma/dy_{lab}\rangle_{|y_{lab}|<1}$ and $T=E_{T_0}/n$
for $\rho^0$, $K^{*0}$ and $\phi$ mesons from a fit of 
function~(\ref{pawer}) to the differential cross sections. 
The average transverse energy $\langle E_T\rangle$, kinetic energy 
$\langle E_T^{kin}\rangle$ and momentum $\langle p_T\rangle$ are also presented.
The errors correspond to the quadratically summed statistical and systematic errors. 
Also shown are measurements in $pp$ and Au-Au
interactions at nucleon-nucleon centre-of-mass energy
$\sqrt{s_{NN}}=200$ GeV~\cite{STAR} at central rapidities.
}
\end{table}
\begin{table}
    \begin{center}
     \begin{tabular}{|l|l|c|}
      \hline
      Experiment &Measurement & $R(\phi/K^{*0})$\\ 
      \hline
       H1&$\gamma p$, $\langle W\rangle=210$~GeV, $|y_{lab}|<1$ & $0.354\pm0.060$\\
      \hline
       STAR&$pp$, $\sqrt{s}=200$~GeV, $|y|<0.5$ & $0.35\pm0.05$\\
      \cline{2-3}
       &Au-Au, $\sqrt{s_{NN}}=200$~GeV, $|y|<0.5$ & $0.63\pm0.15$\\ 
      \hline
     \end{tabular}
    \end{center}
\caption{The ratio $R(\phi/K^{*0})$ of the total cross-sections 
for $\phi$ and $K^{*0}$ production obtained
in $\gamma p$ collisions (H1) at $\langle W\rangle=210$~GeV.
The errors correspond to the quadratically summed statistical and systematic errors.
Also shown are measurements in $pp$ and Au-Au
interactions at nucleon-nucleon centre-of-mass energy
$\sqrt{s_{NN}}=200$ GeV~\cite{STAR} at central rapidities.
}
\end{table}
\clearpage
\vspace*{4.0cm}

\clearpage
\vspace*{4.0cm}
\clearpage

\clearpage


\end{document}